\documentclass[twocolumn,showpacs,preprintnumbers,amsmath,amssymb,prb]{revtex4}

\usepackage{graphicx}
\usepackage{epsf}
\begin{document}

\title{High-frequency transport in $p$-type Si/Si$_{0.87}$Ge$_{0.13}$ heterostructures studied
with surface acoustic waves in the quantum Hall regime}

\author{I. L. Drichko}
\email{irina.l.drichko@mail.ioffe.ru}
\author{A. M. Diakonov}
\author{I. Yu. Smirnov}
%\email{ivan.smirnov@mail.ioffe.ru}
\author{G. O. Andrianov}
\affiliation{A. F. Ioffe Physico-Technical Institute of Russian
Academy of Sciences, 194021 St.Petersburg, Russia}

\author{O. A. Mironov}
\author{M. Myronov}
\author{D. R. Leadley}
\author{T. E. Whall}
\affiliation{Department of Physics, University of Warwick,
Coventry CV4 7AL, UK}

\date{\today}

\begin{abstract}
{The interaction of surface acoustic waves (SAW) with $p$-type
Si/Si$_{0.87}$Ge$_{0.13}$ heterostructures has been studied for
SAW frequencies of 30-300 MHz.  For temperatures in the range
0.7$<T<$1.6 K and magnetic fields up to 7 T, the SAW attenuation
coefficient $\Gamma$ and velocity change $\Delta V /V$ were found
to oscillate with filling factor.  Both the real $\sigma_1$  and
imaginary $\sigma_2$ components of the high-frequency conductivity
have been determined and compared with quasi-dc magnetoresistance
measurements at temperatures down to 33 mK.  By analyzing the
ratio of $\sigma_1$ to $\sigma_2$, carrier localization can be
followed as a function of temperature and magnetic field. At
$T$=0.7 K, the variations of $\Gamma$, $\Delta V /V$ and
$\sigma_1$ with SAW intensity have been studied and can be
explained by heating of the two dimensional hole gas by the SAW
electric field. Energy relaxation is found to be dominated by
acoustic phonon deformation potential scattering with weak
screening.}
\end{abstract}


\maketitle

\section{Introduction}
\label{Introduction}

As is well known,~\cite{book} in the quantum Hall effect (QHE)
regime the magnetic field dependences of the off-diagonal
$\sigma_{xy}^{DC}$ and diagonal $\sigma_{xx}^{DC}$ components of
the dc conductivity tensor are very different. Namely,
$\sigma_{xy}^{DC}$ shows a set of plateaus with abrupt steps
taking place at \emph{half-integer} values of the filling factor
$\nu$. By contrast, $\sigma_{xx}^{DC}$ is extremely small at the
quantum Hall plateaux and has sharp maxima at the steps between
the plateaux. The conventional explanation is that, at a
half-integer filling, electronic states at the Fermi level are
\emph{extended} while at all other filling factors they are
\emph{localized}.

A powerful way to investigate two-dimensional systems, and
effectively probe heterostructure parameters, is with a surface
acoustic wave (SAW),~\cite{gen_ref} especially as this is a
\emph{non-contact} measurement: it does not require a Hall-bar to
be configured and there is no need for carrier injection at the
low-dimensional interface.  Moreover, simultaneous measurement of
the attenuation and velocity of the SAW provides a unique way to
determine the \emph{complex} ac conductivity
$\sigma_{xx}(\omega)=\sigma_1(\omega)-i \sigma_2(\omega)$ as a
function of magnetic field, temperature $T$ and SAW frequency
$\omega$. Furthermore, the magnetic field dependence of
$\sigma_{xx}(\omega)$ provides information on both the extended
and localized states and allows an analysis of the interplay
between these states.~\cite{ildPRB}

As we observed earlier in GaAs/AlGaAs heterostructures,~\cite{1,2}
near steps in the Hall conductance, i.e. at half-integer $\nu$,
the imaginary part of the complex ac conductance $\sigma_2
(\omega)$ is small while the real part $\sigma_1 (\omega)$
coincides with the dc transverse conductance $\sigma_{xx}^{DC}$.
However, in magnetic fields corresponding to regions near the
center of the Hall plateaux, i.e. at small integer $\nu$, the
difference between $\sigma_{xx}(\omega)$ and $\sigma_{xx}^{DC}$
turns out to be crucial. Namely, $\sigma_{xx}^{DC}$ is extremely
small while both $\sigma_1 (\omega)$ and $\sigma_2 (\omega)$ are
measurable quantities and $\sigma_2 (\omega) \gg  \sigma_1
(\omega)$. According to Ref.~\onlinecite{Efros} these facts lead
to the conclusion that the mechanism of ac conductance is
\emph{hopping}.

In the present work, for the first time, an acoustic method has
been applied in a study of $p$-type Si/Si$_{0.87}$Ge$_{0.13}$
heterostructures with the purpose of clarifying high-frequency
transport mechanisms in the system.  Since Ge and Si are not
piezoelectric the only way to measure acousto-electric effects in
these structures is via a hybrid method: a SAW propagates along
the surface of piezoelectric LiNbO$_3$  while the Si/SiGe sample
is gently pressed onto the LiNbO$_3$ surface by means of a spring.
In this case strain from the SAW is not transmitted to the sample
and it is only the electric field accompanying the SAW that
penetrates the sample and creates currents that, in turn, produce
feedback to the SAW.  As a result, both the SAW attenuation
coefficient $\Gamma$  and velocity $V$ appear to depend on the
properties of the two-dimensional hole gas (2DHG).

In this paper we present experimental results for SAW frequencies
up to 300 MHz, consider localization effects and analyze the
non-linear SAW interaction with the 2DHG.

\section{Experiment and discussion}

\label{Experiment and discussion}

\begin{figure}[h]
\centerline{
\includegraphics[width=6cm,clip=]{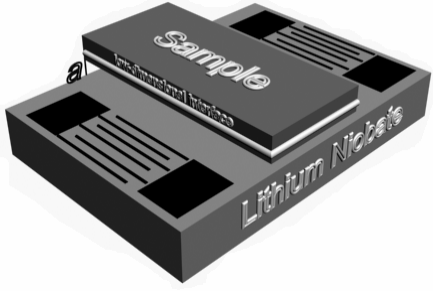}}
\caption{Schematic diagram of the acoustoelectric device.}
\label{setup}
\end{figure}

\begin{figure}[h] \centerline{
\includegraphics[width=5cm,clip=]{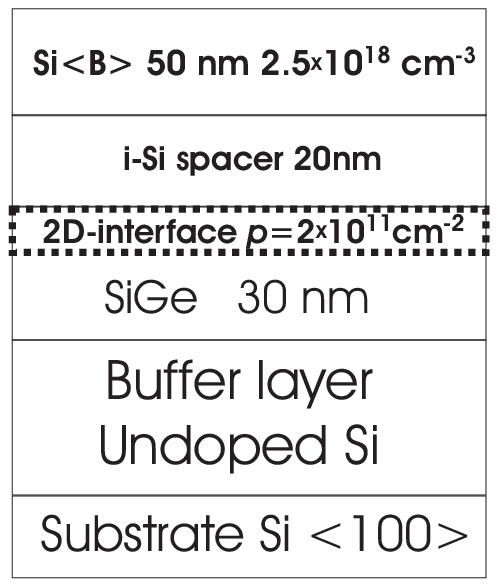}}
\caption{Layer scheme of the sample.} \label{sample}
\end{figure}

In our experimental arrangement, shown schematically in
Fig.~\ref{setup}, the SAW is induced by interdigital transducers
at the surface of a piezoelectric LiNbO$_3$ plate on top of which
the sample has been placed.  The samples were modulation doped
Si/Si$_{0.87}$Ge$_{0.13}$ heterostructures with a 2DHG sheet
density $p = 2\times 10^{11}$ cm$^{-2}$ and a mobility $\mu =
10500$ cm$^2$/Vs, measured at $T$=4.2 K.  The layered system
(shown in Fig.~\ref{sample}) was grown by molecular beam epitaxy
at Warwick University starting from a Si(100) substrate, and
consisting of an undoped Si buffer layer followed by a 30 nm
Si$_{0.87}$Ge$_{0.13}$ strained quantum well, a 20 nm undoped
spacer and finally a 50 nm boron doped Si layer with an acceptor
concentration of 2.5 $\times 10^{18}$ cm$^{-3}$.\cite{Miron}

\begin{figure}[h] \centerline{
\includegraphics[width=8cm,clip=]{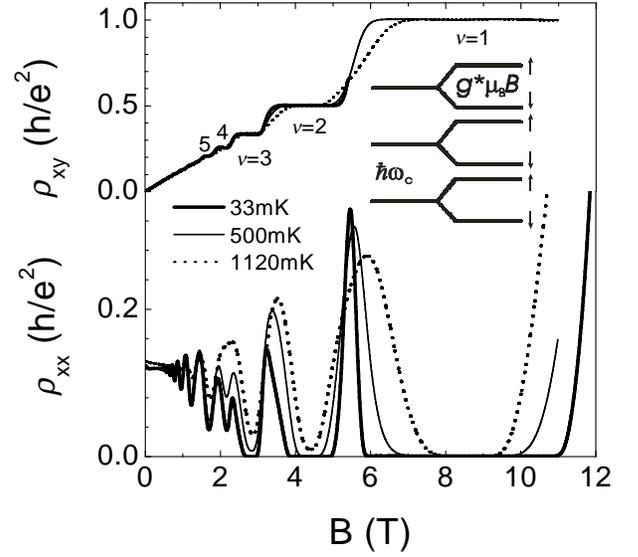}}
\caption{Resistivity components, in units of $h/e^2$ at different
temperatures.} \label{DC}
\end{figure}

Quasi-DC measurements (7-17 Hz) of the resistivity components
$\rho_{xx}$ and $\rho_{xy}$ have been carried out in magnetic
fields up to 11 T in the temperature range 33 mK -1.3 K and show
the integer quantum Hall effect (Fig.~\ref{DC}).  One interesting
feature of transport properties in SiGe 2DHGs is the dominance of
minima associated with odd filling factors for $\nu>$ 2.  This is
because the spin splitting is comparable to the cyclotron energy,
with the enhanced exchange interaction increasing the effective
$g$-factor to 4.5,~\cite{FNTMir,Fang,Glaser,Coleridge} and it is
illustrated in the inset. As can be seen from Fig.~\ref{DC}, the
minima in $\rho_{xx}$ are more pronounced at $\nu$=3 and 5 than at
$\nu$=4, which only appears below 0.5 K. The DC transport
measurements were made on a Hall bar sample as electrical contacts
were not present on the SAW sample. During the experiments the
samples were cooled down several times with only very small
differences in conductivity on each cycle, so whilst the dc
conductivity trace does not completely represent that of the
sample at the time of SAW investigation we believe it gives a very
good indication within experimental uncertainty less than 1$\%$.

\begin{figure}[h]
\centerline{
\includegraphics[width=9cm,clip=]{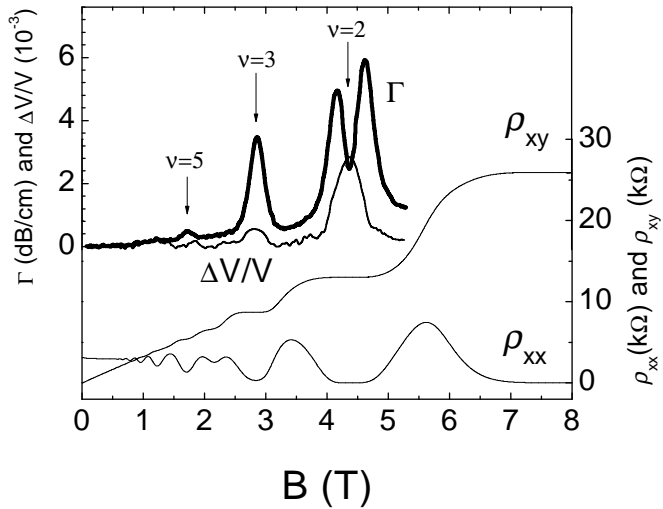}}
\caption{Magnetic field dependence of $\Gamma$ and  $\Delta V /V$
with a SAW frequency of 30MHz together with the dc resistivity
components $\rho_{xx}$ and $\rho_{xy}$. All at T=0.7 K}
\label{GamVel}
\end{figure}

The absorption, or attenuation, coefficient $\Gamma$ and relative
velocity change $\Delta V /V$ of the SAW that interacts with the
2DHG in the SiGe channel, have been simultaneously measured at
temperatures $T$=0.7-1.6 K in perpendicular magnetic fields up to
7 T.  Fig.~\ref{GamVel} illustrates the field dependence of
$\Gamma$ and $\Delta V /V$ for a frequency of 30MHz at 0.7 K
together with the magnetoresistance components. One can see that
the absorption coefficient and the velocity change both undergo
Shubnikov-de Haas (SdH) type oscillations in magnetic field, with
maxima corresponding to $\rho_{xx}$ minima and the centers of the
IQHE plateaus in $\rho_{xy}$, thus allowing the hole density in
the 2D-channel to be determined directly from acoustic
measurements. One exception is the maximum of the attenuation
coefficient in the vicinity of $\nu$=2, which splits to reveal a
minimum in $\Gamma$ at exactly $\nu$=2 where the velocity shift is
at its largest.  This behavior has previously been seen in GaAs
samples ~\cite{gen_ref,ildPRB} and can be understood through the
dependence of $\Gamma$ on conductivity (see Eqs.~\ref{G} and
\ref{V} below).

Simultaneous measurements of $\Gamma$ and  $\Delta V /V$ enable
the high frequency conductivity $\sigma_{xx}(\omega)=
\sigma_{1}-i\sigma_{2}$ to be extracted using the following
equations from Ref.~\onlinecite{ildPRB}:
\begin{eqnarray}
\Gamma=8.68qA(q)\frac{K^2}{2}\frac{\Sigma_1}{\Sigma_1^2
+(1+\Sigma_2)^2}\, , \ \frac{\text{dB}}{\text{cm}}\, \label{G}
\end{eqnarray}
\begin{eqnarray}
\frac{\Delta V}{V}=A(q)\frac{K^2}{2}\frac{1+\Sigma_2  }{\Sigma_1^2
+(1+\Sigma_2)^2}\, ,\label{V}
\end{eqnarray}
where  $A(q)= 8
b(q)(\varepsilon_1+\varepsilon_0)\varepsilon_0^2\varepsilon_s
e^{-2 q(a+d)}$, and $\Sigma_{i}= (4\pi \sigma_{i}/\varepsilon_s
V)\,t(q)$,
\begin{eqnarray}
b(q)=(b_1(q)[b_2(q)-b_3(q)])^{-1}
  \, , \nonumber  \end{eqnarray}
\begin{eqnarray}
t(q)=[b_2(q)-b_3(q)]/2b_1(q)\, , \nonumber \end{eqnarray}
\begin{eqnarray}
b_1(q)=(\varepsilon_1+\varepsilon_0)(\varepsilon_s+\varepsilon_0)
- (\varepsilon_1-\varepsilon_0)
(\varepsilon_s-\varepsilon_0)e^{-2qa}\, ,  \nonumber
\end{eqnarray}
\begin{eqnarray}
b_2(q)=(\varepsilon_1+\varepsilon_0)(\varepsilon_s+\varepsilon_0)
+ (\varepsilon_1+\varepsilon_0)
(\varepsilon_s-\varepsilon_0)e^{-2qd}\, ,  \nonumber
\end{eqnarray}
\begin{eqnarray}
b_3(q)=
(\varepsilon_1-\varepsilon_0)(\varepsilon_s-\varepsilon_0)e^{-2qa}
+ \nonumber \\
+(\varepsilon_1-\varepsilon_0)
(\varepsilon_s+\varepsilon_0)e^{-2q(a+d)} \nonumber.
\end{eqnarray}
Here $K^2$ is the electromechanical coupling constant for lithium
niobate (Y-cut), $q$ is the SAW wave vector; $\varepsilon_1$=50,
$\varepsilon_0$=1 and  $\varepsilon_s$=11.7 are the dielectric
constants of LiNbO$_3$, the vacuum and bulk Si, respectively. In
these calculations we can ignore the small difference of 4 $\%$ in
dielectric constant between Si and Si$_{0.87}$Ge$_{0.13}$. The
value of $d$=70 nm denotes the finite distance between the sample
surface and the 2DHG layer and $a$ is the vacuum clearance between
the sample surface and the LiNbO$_3$ surface.  This clearance
remains finite, despite the heterostructure being pressed to the
piezoelectric platelet, because of some roughness of both
surfaces. Since the actual clearance is not well controlled, the
quantity $a$ is treated as an adjustable parameter.  It is
determined by fitting the experimental data at those magnetic
fields where the conductance is metallic and essentially frequency
independent.  In the present experiments the clearance $a$ ranges
from 0.5 to 1.0 $\mu m$.\cite{1}

Measurements and analysis of the high-frequency conductivity will
be considered in two regimes: (A) where the response is linear in
applied SAW power and (B) the nonlinear region, when acoustic
effects begin to depend on the intensity of the sound wave.

\subsection{Linear regime}

\label{Linear regime}

\begin{figure}[t]
\centerline{
\includegraphics[width=8.5cm,clip=]{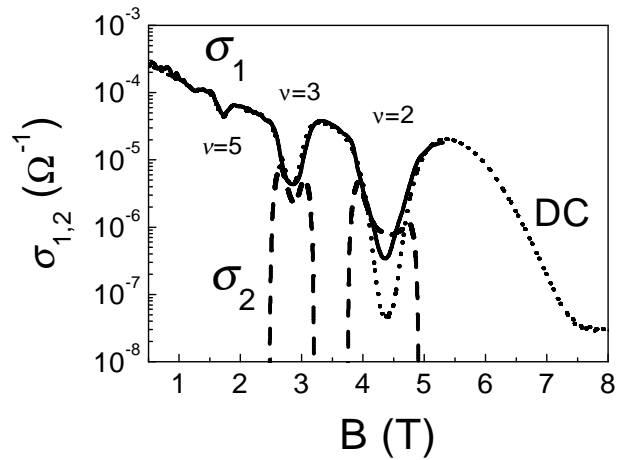}} \caption{Magnetic field
dependence of  $\sigma_{1}$  (solid), $\sigma_{2}$  (dashed) at
$f$=30MHz and $\sigma_{xx}^{DC}$ (dotted); all at T=0.7 K.}
\label{S12}
\end{figure}

Fig.~\ref{S12} illustrates the magnetic field dependence of the
real $\sigma_{1}$ and imaginary $\sigma_{2}$ components of the
complex high-frequency conductivity, obtained from SAW
measurements and using Eqs.~\ref{G} and \ref{V}, together with the
dc-conductivity, derived from magnetoresistance measurements via
$\sigma_{xx}^{DC}=\rho_{xx}/(\rho_{xx}^2+\rho_{xy}^2)$, all at
$T$=0.7K.  One can see that $\sigma_{1}$ demonstrates SdH-type
oscillations and, in the same way as $\rho_{xx}^{DC}$, is
dominated by features at odd filling factors - the only minimum at
even occupancy is observed at $\nu$=2.

It is also evident from Fig.~\ref{S12} that $\sigma_{1}$ is equal
to the static conductivity $\sigma_{xx}^{DC}$ in the low magnetic
field region (below 2 T) and near half-integer filling factors
where the electron states are \emph{delocalized}.  However, at
magnetic fields close to $\nu$=2, where states are strongly
localized, $\sigma_{1} \gg \sigma_{xx}^{DC}$.  Fig.~\ref{S12} also
shows that whilst $\sigma_{2}$ is very small near half-integer
filling factors and at low magnetic fields, it has appreciable
magnitude near to $\nu$=2 and 3. It is actually double
$\sigma_{1}$ at $\nu$=2, which, along with the fact that
$\sigma_{1} \gg \sigma_{xx}^{DC}$, points to a manifestation of
two-dimensional high-frequency hopping conductivity.  It is known
that in the IQHE regime carriers can be localized in the random
potential generated by charged impurities, with the result that
the dc conductivity vanishes and the high-frequency conductivity
has to be via \emph{hopping}.  According to
Ref.~\onlinecite{Efros}, when the 2D high-frequency conductivity
has a hopping character, and is determined by hole hops between
states close in energy but localized at two different impurity
centers (that form \emph{pair complexes} which do not overlap),
$\sigma_{2} \gg \sigma_{1}$ and $\sigma_{xx}^{DC}$=0. The features
we observe at small $\nu$ thus provide evidence that hole hopping
begins to make an important contribution to the high-frequency
conductivity. However, this would appear not to be the case at
lower magnetic fields where $\sigma_{1} = \sigma_{xx}^{DC}$ and
$\sigma_{2}$ vanishes.  In fact at $\nu$=3 $\sigma_{1} =
\sigma_{xx}^{DC} > \sigma_{2}$ so even here hopping is not the
major high frequency conduction mechanism.

In Ref.~\onlinecite{ildPRB} it has been shown that for total
localization the ratio $\sigma_2 /\sigma_1$ will have the form:
\begin{equation}
\frac{\sigma_2}{\sigma_1}=\frac{2{\cal L}_\omega ( {\cal L}_T^2 +
{\cal L}_T {\cal L}_\omega/2 + {\cal L}_\omega^2/12 )+ 4c {\cal
L}_T^2\, {\cal L}_c}{\pi ( {\cal L}_T^2 + {\cal L}_T {\cal
L}_\omega + {\cal L}_\omega^2/4 )}\, , \label{rat1}
\end{equation}
where  ${\cal L}_T =\ln {(J/kT)}$, with $J$ of the order of the
Bohr energy, ${\cal L}_\omega= \ln(G_0/\omega)$, ${\cal L}_c=\ln(
\hbar \omega_c/k_BT)$ and $c \gtrsim 1$ is a numerical factor
depending on the density of states in the region between the
Landau levels. An estimation of $G_0$ valid for the deformation
phonon relaxation mechanism is given in Ref.~\onlinecite{GalpHop}
as $G_0=(k_B T)^3 D_{ac}^2 / \rho \hbar^4 s^5 $, where $D_{ac}$ is
the deformation potential, $\rho$ is the mass density, $s$ is the
longitudinal sound velocity. Using this estimation one concludes
that in the hopping regime $\sigma_2 / \sigma_1 \cong$ 5, whereas
experimentally we observe  $\sigma_2 / \sigma_1 \cong$ 2 in the
middle of the $\nu$=2 Hall plateau. This experimental result
suggests that the high frequency conductivity has a \emph{mixed}
mechanism consisting of two contributions:  The first we assign to
the extended states, while the second, hopping, is associated with
the localized states.  For $\nu$=3, at the lower magnetic field of
2.9 T, the hopping contribution becomes even smaller, and at 1.7 T
($\nu$=5) it is completely absent.

\begin{figure}[t]
\centerline{
\includegraphics[width=6cm,clip=]{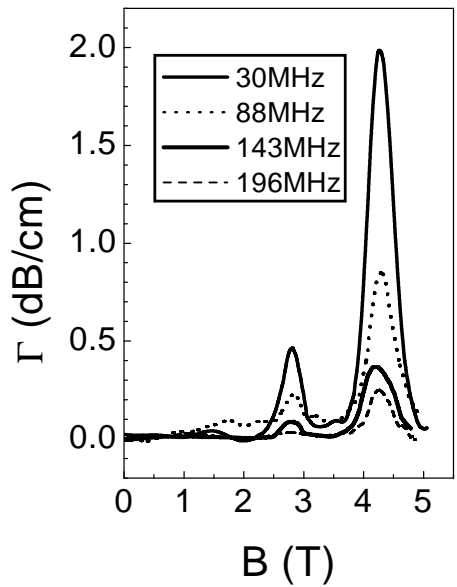}} \caption{
Magnetic field dependences of the absorption coefficient $\Gamma$
at different frequences at $T$=1 K.} \label{GamFreq}
\end{figure}

Fig.~\ref{GamFreq} illustrates the effect on the absorption
coefficient $\Gamma$ of varying the SAW frequency at the
temperature $T \cong$ 1 K.  At this temperature the velocity
change $\Delta V / V$ is small and does not exceed
3$\times$10$^{-4}$ at $\nu$=2. Although there are large changes in
the magnitude of $\Gamma$, these are completely accounted for by
the explicitly frequency-dependent factors of Equation 1.
Therefore, within our experimental error, we find no frequency
dependence of the acoustically measured conductivity
$\sigma_{xx}(\omega)$ at $T \cong$ 1 K. It is somewhat surprising
to see no effect of frequency in the region of  $\nu$=2 where a
significant hopping contribution was found at 0.7 K, but we must
conclude that the increase in temperature has reduced the relative
importance of hopping. Hence, one can see that the acoustic
measurements allow an analysis of the $\sigma_1$ to $\sigma_2$
ratio and, thus, the degree of carrier localization can be
followed as a function of temperature and magnetic field.

\subsection{Nonlinear regime}

\label{Nonlinear regime}

\begin{figure}[h] \centerline{
\includegraphics[width=8cm,clip=]{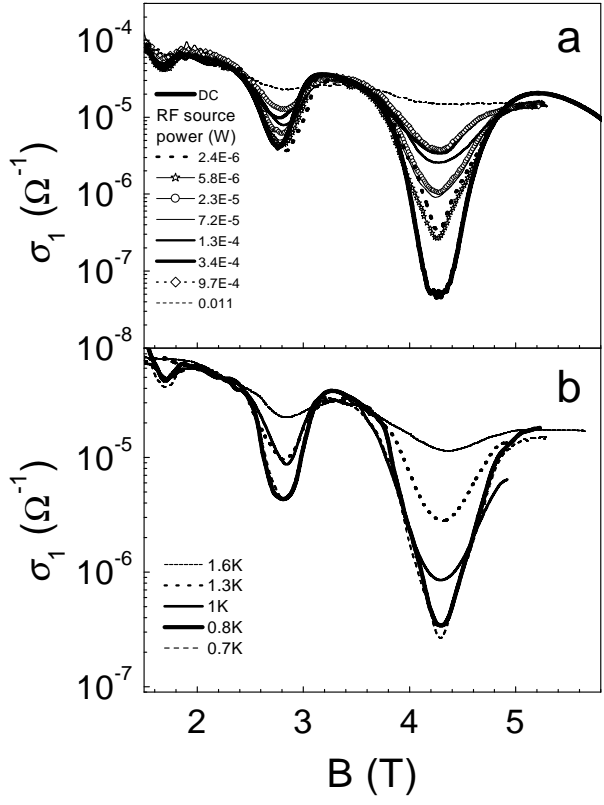}}
\caption{ (a) Dependences of $\sigma_{1}$ on magnetic field $B$ at
different RF-source powers, $T$=0.7 K, $f$=30 MHz; (b)
$\sigma_{1}$ versus $B$ in the linear regime ($P \simeq 5\times
10^{-6}$ W) at different temperatures, $f$=30 MHz.} \label{S1HPT}
\end{figure}

\begin{figure}[h] \centerline{
\includegraphics[width=8.5cm,clip=]{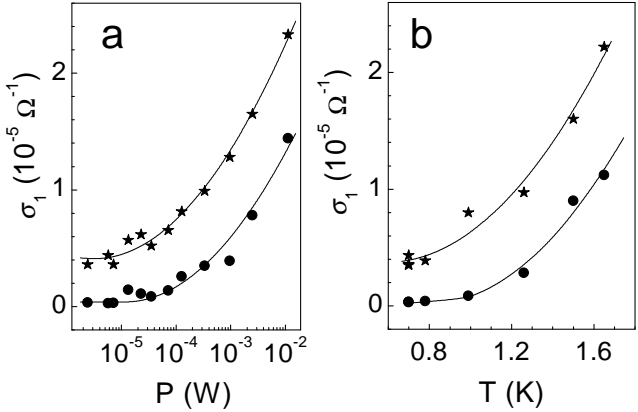}}
\caption{ (a) Dependence of $\sigma_{1}$ on the RF-source power
$P$; (b) temperature variation of $\sigma_{1}$ in the linear
regime.  $f$=30 MHz, Magnetic fields of 4.3 T (circles) and 2.9 T
(stars) correspond to $\nu$=2 and 3 respectively.  The lines are
guides to the eye.} \label{S1PT}
\end{figure}

To investigate non-equilibrium effects we have measured the
dependence of $\Gamma$ and $\Delta V /V$ on the SAW intensity at
30 MHz, by varying the SAW-source power over the range 0.002 - 11
mW at $T$=0.7K.  From these measurements $\sigma_{1}$ has been
extracted and is shown in Fig.~\ref{S1HPT}a for the range of
RF-source powers. It can be seen that $\sigma_{1}$ increases at
higher SAW power and we find that the condition $\sigma_{1} \gg
\sigma_{2}$ is satisfied in this non-linear regime.  We interpret
this to mean that the high-frequency conductivity is mainly from
delocalized holes. These results should be compared with the
magnetic field dependence of $\sigma_{1}$ extracted from
measurements at different temperatures, and sufficiently low power
($<10^{-5}$ W) to remain in the linear regime, illustrated in
Fig.~\ref{S1HPT}b. For a quantitative comparison, values of
$\sigma_{1}$ have been taken from Fig.~\ref{S1HPT} at magnetic
fields of 2.9 T and 4.3 T, corresponding to $\nu$=3 and $\nu$=2
respectively, and plotted in Fig.~\ref{S1PT} as a function of (a)
RF-source power $P$ and (b) temperature.  One can see from these
plots that $\sigma_{1}$ shows similar increases with temperature
and SAW power leading to the conclusion that the observed
nonlinear effects are probably associated with carrier heating in
the SAW electric field.

There have been a number results published on hole heating by a
steady electric field in the 2DHG of Si/SiGe
heterostructures.~\cite{MirAgan,MirDeform,Leturcq} However, it is
valuable to investigate the energy-loss mechanisms in these
structures using a contactless method (such as possible here using
acoustics) that excludes carrier injection to the low-dimensional
interface from contact areas.  Hole heating may be described by
means of a carrier temperature $T_c$, greater than the lattice
temperature $T$. (The propriety of introducing this carrier
temperature will be discussed below).  From Fig.~\ref{S1PT} $T_c$
can be determined by comparing the variation of $\sigma_{1}$ on
SAW power $P$ with the curves of $\sigma_{1}$ versus the lattice
temperature $T$, in a manner analogous to SdH
thermometry.~\cite{Leadley,HeatSAW} Such a comparison makes it
possible to establish a correspondence between the temperature of
the 2DHG and the RF-source power.

\begin{figure}[h] \centerline{
\includegraphics[width=7.0cm,clip=]{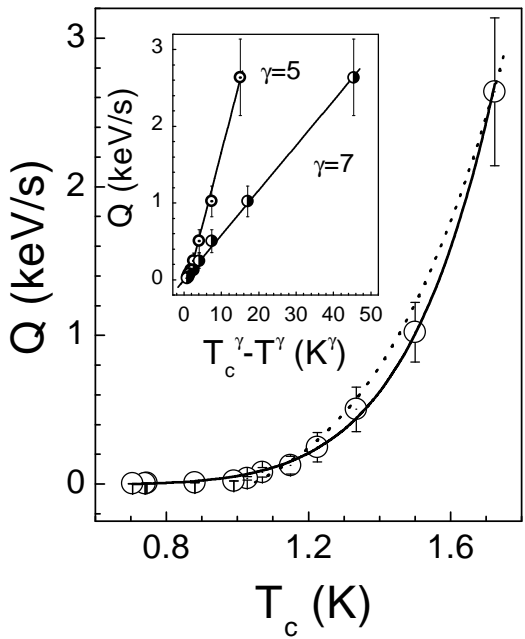}}
\caption{The energy losses rate per hole measured in acoustics
$Q=\bar{Q}/p$ plotted versus $T_c$. Lines are the fitting curves
with $\gamma=5$ (dotted) and $\gamma=7$ (solid). Inset: dependence
of $Q$ on ($T_c^{\gamma}-T^{\gamma}$). Lines are results of the
linear fit.} \label{QT}
\end{figure}

To characterize the energy relaxation mechanisms one needs to
extract the absolute energy loss rate $\bar{Q}$ resulting from the
SAW interaction with the carriers.  Since only delocalized states
are being considered and $\sigma_{2}$=0 we can use the result of
Ref.~\onlinecite{HeatSAW} that the energy loss rate is
$\bar{Q}=\sigma_{xx}E^2=4\Gamma W$, where $E$ is the SAW electric
field,

\begin{eqnarray}
\ |E|^2=K^2\frac{32\pi}{V}
\frac{(\varepsilon_1+\varepsilon_0)\varepsilon_0^2qe^{(-2q(a+d))}}
{b_1^2(q)[1+(\frac{4\pi \sigma_{xx}(\omega)}{\varepsilon_s
V}t(q))^2]}W,\label{eq2}
\end{eqnarray}
and $W$ is the input SAW power scaled to the width of the sound
track.  The resulting energy losses rate per hole $Q=\bar{Q}/p$
are shown as a function of $T_c$ in Fig.~\ref{QT}.

From numerous works on dc heating \cite{MirAgan,MirDeform,Leturcq}
a consensus appears that the energy relaxation of holes in Si/SiGe
2DHGs is due to acoustic phonon deformation potential scattering.
However, the authors disagree over whether a screened or
unscreened deformation potential should be used. The functional
form $Q=A_{\gamma} (T_c^{\gamma}-T^{\gamma})$, corresponds to
energy relaxation via carrier scattering from the deformation
potential of the acoustic phonons with either weak screening
($\gamma$=5) or strong screening ($\gamma$=7) and the slope of
inset of Fig.~\ref{QT} is either: \cite{Karpus}
\begin{eqnarray}
A_5=\frac{3 \sqrt{2} m^{*2} \zeta(5) D^2_{ac} k^5_B}{\pi^{5/2} s^4
\hbar^7 p^{3/2} \rho}, \label{A5}
\end{eqnarray}
or: \cite{YMa, Price, Pudalov}
\begin{eqnarray}
A_7=\frac{45\sqrt{2} m^{*2} \zeta(7) D^2_{ac} k^7_B}{ \pi^{5/2}
s^6 \hbar^9 p^{3/2} \rho q_s^2}, \label{A7}
\end{eqnarray}
where $m^*$=0.24$m_e$ is the effective mass for
Si$_{0.87}$Ge$_{0.13}$ \cite{FNTMir} and $q_s=2 m^*
e^2/\varepsilon_s \hbar^2 =7.8 \times 10^6$ cm$^{-1}$ is the
screening wavevector for $p$-type SiGe.  A least-squares analysis
showed that the experimental curves of Fig.~\ref{QT} could be
fitted by the functional form $Q=A_5 (T_c^5-T^5)$, but given the
small temperature range of the data the form $Q=A_7 (T_c^7-T^7)$
is equally applicable. In the inset of Fig.~\ref{QT}, $Q$ is
plotted against $ (T_c^\gamma-T^\gamma)$ for both $\gamma$=5 and 7
revealing a linear dependence in each case. According to
Ref.~\onlinecite{Pol}, screening of the hole-phonon interaction
can be neglected if the screening wavevector $q_s$ exceeds the
wavevector $q_{\varepsilon}$ of phonons with the average thermal
energy, which in our case is $q_{\varepsilon}$=8.2$\times$10$^5$
cm$^{-1}$.  Thus, $q_s \gg q_{\varepsilon}$ and the energy
relaxation here is dominated unambiguously by \emph{unscreened}
acoustic phonon scattering so we should use $\gamma$=5. This is in
agreement with previous work on pseuodomorphic Si$_{1-x}$Ge$_x$
heterostructures with x=0.2 \cite{MirDeform} and x=0.15
\cite{Leturcq}, where weak screening was found to be appropriate
for a wide range of hole density up to at least 10$^{12}$
cm$^{-2}$. From the expression for $A_5$, the value of the
deformation potential can be determined as $D_{ac}$=5.3$\pm$0.3
eV.  This value is in good agreement with the deformation
potential estimate of 5.5$\pm$0.5 eV obtained in
Ref.~\onlinecite{MirAgan} for the same SiGe heterostructure from
phonon-drag thermopower measurements and, although not agreeing,
is of similar magnitude to the values of 3$\pm$0.4 eV and
2.7$\pm$0.3 eV reported in Refs. \onlinecite{MirDeform} and
\onlinecite{Leturcq} respectively.

\section{Conclusion}

\label{Conclusion}

For the first time, the \emph{contactless} acoustic method has
been applied to strained $p$-type Si/SiGe heterostructures.
Simultaneous measurements of the attenuation and velocity shift of
a surface acoustic wave were conducted and used to obtain the high
frequency conductivity and its dependence on magnetic field,
temperature and SAW power. The following conclusions may be drawn
from our experimental results and analysis:

In the integer quantum Hall effect regime, the degree of the
carrier localization can be followed as a function of magnetic
field and/or temperature from the high frequency conductivity.  It
is found that even at $T$=0.7 K close to the center of the $\nu$=2
Hall plateau, only some fraction of the holes in the 2D-channel
appear to be localized.  Thus, the mechanism responsible for the
high frequency conductivity appears to be a mixture of two terms
and is determined both by the extended states as well as hopping
via localized states.

In studying non-linear effects, heating of the 2DHG in Si/SiGe
heterostructures by the SAW electric field is observed.  This
heating could be described by a carrier temperature $T_c > T$.
From the experimental dependence of the energy loss rate on $T_c$,
we find that the hole energy relaxation time $\tau_{\varepsilon}$
is determined by dissipation in the unscreened acoustic phonon
deformation potential  and have obtained values for
$\tau_{\varepsilon}$ as well as the deformation potential.

\acknowledgments The work was supported by RFFI 04-02-16246,
NATO-CLG 979355, Grant of the President of the RF NS-2200.2003.2,
Prg. MinNauki "Spintronika" and INTAS-01-0184.

\appendix

\vspace{12pt}

\section{Justification for using $T_c$} \label{Applicability conditions}
The condition for introducing a carrier temperature $T_c > T$ is
that relaxation processes within the carrier gas must be much
faster than those leading to thermalization of the carriers i.e.

\begin{eqnarray}
\tau_0 << \tau_{cc} << \tau_{\varepsilon}, \label{taurel}
\end{eqnarray}
where $\tau_0$, $\tau_{cc}$, and $\tau_{\varepsilon}$ are the
momentum, carrier-carrier, and energy relaxation times,
respectively.  The momentum relaxation time can be found from the
transport mobility as $\tau_0=\mu_0 m^*/e=1.4 \times 10^{-12}
~\text{s}$ and is clearly shorter than the carrier-carrier
interaction time given by \cite{1,FNTMir}
\begin{eqnarray}
\tau_{cc}=\frac{k_B T \rho_{xx} e^2}{2 \pi \hbar^2} \ln \frac{h}{2
e^2 \rho_{xx}}= 6.4 \times 10^{-11} ~\text{s}. \label{taucc}
\end{eqnarray}

Finally, we can estimate the energy relaxation time
$\tau_{\varepsilon}$ from our acoustic measurements.  Indeed, if
the 2D gas can be characterized by a carrier temperature $T_c$ and
heating is weak $\Delta T =T_c - T \ll T$, the energy loss rate
per hole can be written in the form \cite{Gantmakher}
$Q=[\bar{\varepsilon}(T_c)- \bar{\varepsilon}(T)]/
\tau_{\varepsilon}$, where $\bar{\varepsilon}(T_c)$ and
$\bar{\varepsilon}(T)$ are the average carrier energy at $T_c$ and
$T$, respectively.  Assuming also that $\varepsilon_F \gg k_BT$,
the change in average kinetic energy of a hole
$\Delta\varepsilon=\bar{\varepsilon}(T_c)- \bar{\varepsilon}(T)$
is $\Delta\varepsilon=\frac{\pi^2k_B^2}{3} \frac{T\Delta
T}{\varepsilon_{F}}$. Then, by expanding our previous expression
for the loss rate $Q=A_{\gamma} (T_c^{\gamma}-T^{\gamma})$ for
small $\Delta T/T$ as $Q=\gamma A_{\gamma} T^{\gamma -1}\Delta T$,
we obtain \cite{HeatSAW}
\begin{eqnarray}
\tau_{\varepsilon}=\frac{(\pi k_B)^2}{3\gamma A_{\gamma}
\varepsilon_F T^{\gamma -2}}. \label{taue}
\end{eqnarray}
Using this equation with $\gamma$=5 and finding $A_{\gamma}$ from
the slope of inset of Fig.~\ref{QT}, we have computed
$\tau_{\varepsilon} =(3.8 \pm 0.4) \times 10^{-8}$ s.  Thus,
relation (\ref{taurel}) is satisfied and it is appropriate to
introduce a carrier temperature $T_c$.

\end{document}